\documentclass[epjCONF, onecolumn]{svjour}
\usepackage{graphics}
\usepackage[varg]{txfonts} % Times fonts
\usepackage[latin1]{inputenc}
\session-title{Assembling the puzzle of the Milky Way}
\begin{document}
\title{Dynamical evolution of a bulge in an N-body model of the Milky Way}
\author{Kanak Saha\inst{1}\fnmsep\thanks{\email{saha@mpe.mpg.de}} \and Inma Martinez-Valpuesta \inst{1} \and Ortwin Gerhard \inst{1} }
\institute{MPE, Garching, Germany}%\and MPE, Garching, Germany \and MPE, Garching, Germany}
\abstract{
The detailed dynamical structure of the bulge in the Milky Way is currently
 under debate. Although kinematics of the bulge stars can be well reproduced 
by a boxy-bulge, the possible existence of a small embedded classical bulge 
can not be ruled out. We study the dynamical evolution of a small classical bulge
in a model of the Milky Way using a self-consistent high resolution 
N-body simulation. Detailed kinematics and dynamical properties of such a bulge
 are presented.} 

%end of abstract
%
\maketitle
\section{Introduction}
\label{intro}
\vspace{-0.2cm}
In the standard $\Lambda$CDM cosmology, nearly non-rotating classical bulges which are the central 
building blocks in spiral galaxies, are generally formed in dry major mergers \cite{hopkins10}. 
As mergers were nearly inescapable in the past and $\sim 2/3$ of the disk 
galaxies \cite{Menendez07} are barred including the Milky Way, the possible co-existence of a bar 
and a small classical bulge might be rather common in present day disk galaxies, and such bulges 
might have evolved through their mutual gravitational interaction. In the Milky Way, an upper 
limit on the mass of a classical bulge ($\sim 8 \%$ of the disk mass) has been set by modelling 
the kinematics from the Bulge Radial Velocity Assay (BRAVA) data \cite{shen10}. But there is 
evidence for a metallicity gradient above the Galactic plane \cite{zoccali08} which 
is taken as an indication for the existence of a classical bulge in our Galaxy. Hence, it is 
important to understand the dynamical interaction between a preexisting classical bulge and the 
bar in the Galaxy. 

\begin{figure}[ht]
\resizebox{0.95\columnwidth}{!}{
\includegraphics{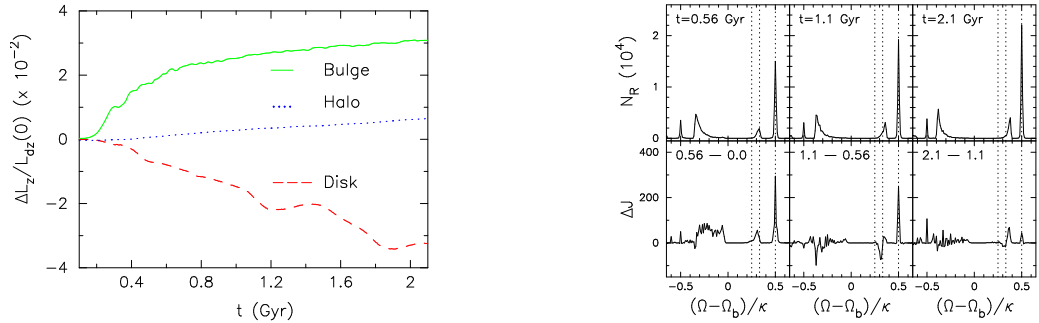}}
\caption{Left: Evolution of the change in the specific angular momentum normalized to disk angular 
momentum at T=0. Right: Spectral analysis showing 2:1 resonance mainly responsible for transferring 
angular momentum to the classical bulge from the bar rotating with a pattern speed $\Omega_b$. 
Other notations bear usual meaning.}
\label{fig:Lexch}       
\end{figure}
\vspace{-0.3cm}
\section{Evolution of a classical bulge: structure and kinematics}\label{sec:clb}
In order to follow the dynamical evolution of a small, initially isotropic, non-rotating 
classical bulge, we construct an equilibrium model of a live disk galaxy consisting of 10 
million particles using the method of \cite{KD95}. The disk density follows an exponential 
profile with a scale length of $4$ kpc, total mass $M_d = 4.5 \times 10^{10} M_{\odot}$
and $Q=1.4$ at half-mass radii. The bulge-to-disk mass ratio (B/D) is 0.067. Other details 
of the galaxy model can be found in \cite{sahaetal11}.  

A strong bar forms in the disk within 0.5 Gyr and is transformed into a boxy bulge
via the well-known buckling instability. During the secular evolution that bar drives a 
substantial fraction of the energy and angular momentum from the disk are being transferred 
to the surrounding dark matter halo and the preexisting classical bulge. Based on the work 
of Lynden-Bell \& Kalnajs \cite{LBK72}, several authors have emphasized that resonant 
interaction plays a significant role in the angular momentum transfer. The left panel of 
Fig.~\ref{fig:Lexch} shows the evolution of specific angular momentum gained by the 
embedded classical bulge. Orbital spectral analysis reveals 
that the 2:1 resonance (right panel of Fig.~\ref{fig:Lexch}) plays the dominant role in the 
transfer of angular momentum from the bar to the classical bulge \cite{sahaetal11}. As a result 
of the angular momentum gain, the initially non-rotating isotropic low mass classical bulge 
transforms into a highly rotating triaxial and anisotropic object. 
\begin{figure}[t]
\centering
\resizebox{0.92\columnwidth}{!}{
\includegraphics{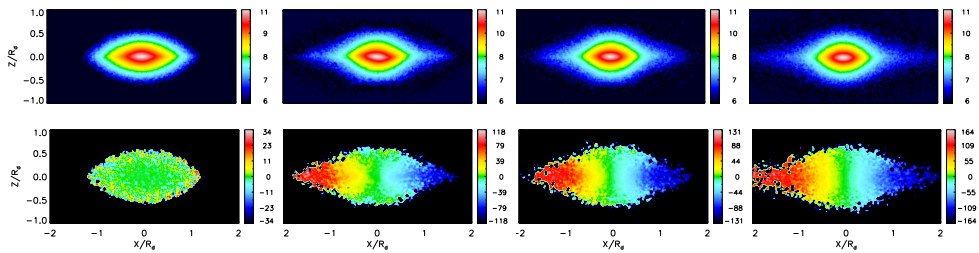}}
\caption{Surface density and velocity maps of the classical bulge alone. From left to right, panels are taken at T=0, 0.56, 1.1 and 2.1 Gyr. The bulge is initially non-rotating and flattened by the disk gravity.}
\label{fig:kinema}
\end{figure}

The angular momentum gained by the low mass classical bulge has a profound effect on its 
structure, kinematics and dynamics. In the upper panel of Fig.~\ref{fig:kinema}, we show
edge-on surface density maps at four different epochs during the secular evolution in the galaxy. 
The normalized fourth-order Fourier cosine coefficient obtained by analyzing the density field 
indicates the presence of non-axisymmetric features developing inside the classical bulge at 
around $T = 0.56$ Gyr which also marks the bar buckling instability in the disk. At later phases 
of evolution, the inner regions of the classical bulge become rounder and the outer parts become 
disky. The corresponding velocity maps on the lower panel of Fig.~\ref{fig:kinema} show that as the 
bar evolves, the bulge spins up and shows prominent signatures of cylindrical rotation in the 
inner regions (at $X/R_d < 0.4$). Further analysis (see \cite{sahaetal11}) have shown that a 
bar-like structure forms in the inner regions of the classical bulge which may be responsible 
for making the inner regions rounder through heating \cite{sahaetal10} and cylindrical rotation 
seen in the classical bulge. Since cylindrical rotation is considered as a proxy for the boxy 
bulge, our new results on the kinematics of the classical bulge would open up the possibility 
that many boxy bulges might have an underlying small classical bulge in barred galaxies, 
especially the Milky Way.

\end{document}